\documentclass[twocolumn,aps,prl,floatfix,showpacs]{revtex4}
\usepackage{graphicx}
\begin{document}

\title{Paramagnon-induced dispersion anomalies in the cuprates}

\author{R.S. Markiewicz, S. Sahrakorpi, and A. Bansil}
\affiliation{
Physics Department, Northeastern University, Boston MA 02115}
\date{\today}
\begin{abstract}
We report the self-energy associated with RPA magnetic susceptibility in 
the hole-doped Bi$_2$Sr$_2$CuO$_{6}$ (Bi2201) and the electron-doped 
Nd$_{2-x}$Ce$_x$CuO$_4$ (NCCO) in the overdoped regime within the 
framework of a one-band Hubbard model. Strong weight is found in the 
magnetic spectrum around $(\pi ,0)$ at about 360 meV in Bi2201 and 640 meV 
in NCCO, which yields dispersion anomalies in accord with the recently 
observed `waterfall' effects in the cuprates.
\end{abstract}
\pacs{79.60.-i, 71.38.Cn, 74.72.-h, 71.45.Gm}
\maketitle

Very recent angle-resolved photoemission (ARPES) experiments in the 
cuprates have revealed the presence of an intermediate energy scale in the 
300-800 meV range where spectral peaks disperse and broaden rapidly with 
momentum, giving this anomalous dispersion the appearance of a 
`waterfall'\cite{RonK,Ale,Non,Feng,Valla,PanDing}. Similar self-energies 
have also been adduced from optical data\cite{TimCar}. This new energy 
scale is to be contrasted with the well-known low energy `kinks' in the 
50-70 meV range, which have been discussed frequently in the cuprates as 
arising from the bosonic coupling of the electronic system with either 
phonons\cite{pkink} and/or magnetic modes\cite{mkink}. Although low energy 
plasmons\cite{HedLee,WZD} are an obvious choice for the new boson, analysis 
indicates that the plasmons lie at too high an energy of $\sim$1 eV to 
constitute a viable candidate\cite{MBII}. 
Here we demonstrate that paramagnons provide not only an explanation of the 
energy scale but also of the other observed characteristics of the 
waterfall effect in both hole and electron doped cuprates.

For this purpose, we have evaluated the self-energy associated with the 
RPA magnetic susceptibility in the hole-doped Bi$_2$Sr$_2$CuO$_{6}$ 
(Bi2201) and the electron-doped Nd$_{2-x}$Ce$_x$CuO$_4$ (NCCO).\cite{foot6} 
In order to keep the computations 
manageable, the treatment is restricted to the overdoped systems where 
magnetic instabilities are not expected to present a complication. Our 
analysis proceeds within the framework of the one-band Hubbard 
Hamiltonian, where the bare band is fit to the tight-binding LDA 
dispersion\cite{Arun3,foot3}. We incorporate self-consistency by 
calculating the self energy and susceptibility using an approximate 
renormalized one-particle Green function 
\begin{equation}
G=\bar Z/(\omega -\bar\xi_k+i \delta ),
\label{eq:1}
\end{equation}
where $\bar\xi_k=\bar Z(\epsilon_k-\mu)$. 
Here, $\epsilon_k$ are bare energies and $\mu$ is the chemical potential, 
and the renormalization factor is $\bar 
Z\sim (1-\partial\Sigma'/\partial\omega )^{-1}<1$.
The associated magnetic susceptibility is
\begin{equation}
\chi_0(\vec q,\omega) =-\bar Z^2\sum_{\vec k}{\bar f_{\vec k}-\bar f_{\vec k+
\vec q}\over\bar\epsilon_{\vec k}-\bar\epsilon_{\vec k+\vec q}+\omega+i\delta},
\label{eq:2}
\end{equation}
where $\delta$ is a positive infinitesimal, $\bar f_{\vec k}\equiv
f(\bar\epsilon_{\vec k})$ is the Fermi function. The RPA susceptibility is 
given by
\begin{equation}
\chi (\vec q,\omega )={\chi_0(\vec q,\omega )\over
1-U\chi_0(\vec q,\omega )}, 
\label{eq:3}
\end{equation}
with $U$ denoting the Hubbard parameter. 
The self-energy can be obtained straightforwardly from the susceptibility 
via the expression\cite{BrEng} (at $T=0$)
\begin{eqnarray}
\Sigma (\vec k,\omega )={3\over 2}\bar ZU^2\sum_{\vec q}\int_0^{\infty}{d\omega 
'\over\pi}Im\chi (\vec q,\omega ')
\nonumber \\ 
\times\Bigl[{\bar f_{\vec k-\vec q}\over\omega -\bar\xi_{\vec k-\vec q}+\omega '}+
{1-\bar f_{\vec k-\vec q}\over\omega -\bar\xi_{\vec k-\vec q}-\omega '}
\Bigr].
\label{eq:4}
\end{eqnarray}

Concerning technical details, we note that for the generic purposes of 
this study, all computations in this article employ a fixed value $\bar Z$ 
=0.5, which is representative of the band dispersions observed 
experimentally in hole as well as electron doped cuprates.\cite{foot2} 
Self-consistency is then achieved approximately by determining 
values of the chemical potential $\mu$ and the Hubbard parameter $U$ to keep a
fixed doping level and to ensure that the bands are indeed renormalized by the 
average factor $\bar Z =0.5$.  The procedure is relatively simple, but 
it should capture the essential physics of the electron-paramagnon 
interaction, although our treatment neglects the energy\cite{foot4} and momentum 
dependencies of $\bar Z$. Note also that in the overdoped regime considered, 
the effective $U$ values in Bi2201 and NCCO are small enough 
that the system remains paramagnetic and the complications of the 
antiferromagnetic instability are circumvented. Specifically, the 
presented results on Bi2201 are for $x=0.27$ with $\mu=-0.43$~eV and $U=3.2t$, while for 
NCCO, $x=-0.25$ with $\mu=0.18$~eV and $U=4t$.

\begin{figure}  
            \resizebox{8.2cm}{!}{\includegraphics{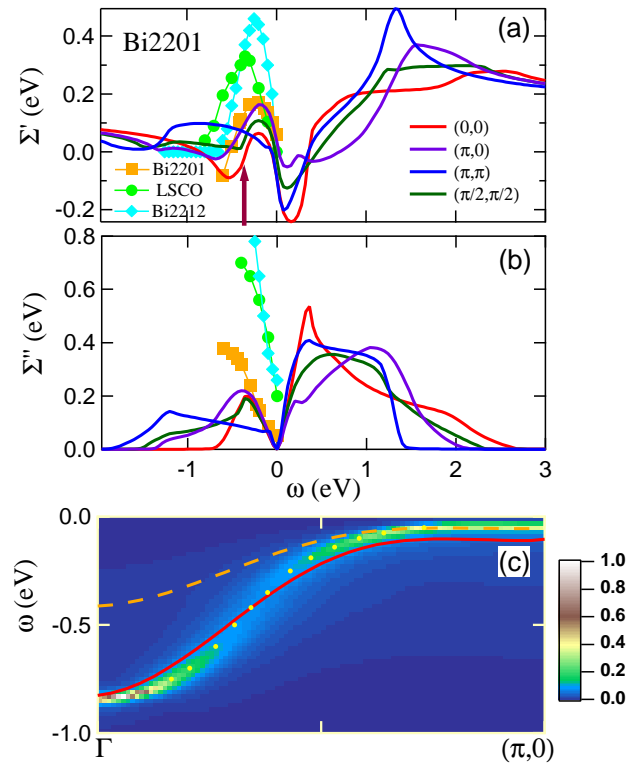}}
\vskip0.5cm
\caption{
(Color online) (a) and (b): Real and imaginary parts $\Sigma'$ and $\Sigma''$, 
respectively, of the paramagnon self-energy in overdoped Bi2201. Theoretical 
results at 
various momenta are shown by lines of different colors. 
The values of $\Sigma'$ have been shifted by a constant, to produce a 
zero average value of the theoretical $\Sigma'$ at the Fermi level.
Experimental 
points are from the nodal point for: Bi2201 (gold squares, 
Ref.~\protect\onlinecite{Non}); LSCO (circles, 
Ref.~\protect\onlinecite{Valla}); and Bi2212 (diamonds, 
Ref.~\protect\onlinecite{Feng}). Thin lines joining experimental points 
are guides to the eye. Arrow marks the location of peak in $\chi''$ at 
$(\pi ,0)$. (c): Spectral density in the energy-momentum plane obtained 
from the dressed Green function is shown in a color plot along with the 
bare (red line) and the renormalized (dashed orange line) dispersions. 
Dots mark the peak positions of the MDC plots of the dressed spectral 
density. 
}
\label{fig:9} 
\end{figure}

Figure~\ref{fig:9} summarizes the results for Bi2201. We consider Figs. 
1(a) and (b) first, which give the real and imaginary parts of the 
self-energy at several different momenta as a function of frequency. The 
theoretical self-energies, which refer to Bi2201, should be compared 
directly with the corresponding experimental data (gold 
squares\cite{Non}), although available experimental points for 
Bi2212\cite{Feng} and La$_{2-x}$Sr$_x$CuO$_4$ (LSCO)\cite{Valla} are also 
included for completeness. 
The agreement between theory and experiment is seen to be quite good for 
the real part of the self-energy in (a), while theory underestimates the 
imaginary part of the self-energy by a factor of $\sim$ 2. That the 
computed $\Sigma''$ is smaller than the experimental one is to be generally 
expected since our calculations do not account for scattering effects 
beyond those of the paramagnons. Here, we should keep in mind that there are 
uncertainties inherent in the experimental self-energies due to different 
assumptions invoked by various authors concerning the bare dispersions in 
analyzing the data. In particular, Feng {\it et al.}\cite{Feng} extract 
the bare dispersion by assuming that $\Sigma '$ is always positive and 
goes to zero at large energies.  Other groups\cite{Non,Ale2} compare their 
results to LDA calculations and argue that $\Sigma '$ must become negative 
at higher energies. Our computed $\Sigma '$ in Fig. 1(a) becomes negative 
over the range 0.35-0.9~eV in certain $\vec k$-directions. Interestingly, 
various computed colored lines in 
(a) and (b) more or less fall on top of one another, indicating that the 
self-energy is relatively insensitive to momentum, especially below the 
Fermi level, consistent with experimental findings\cite{Valla}, even 
though $\Sigma$ possesses a fairly strong frequency dependence.

Fig. 1(c) gives further insight into the nature of the spectral intensity 
obtained from the self energy of Eq.~\ref{eq:4}. The spectral intensity 
shown in the color plot of the figure is representative of the ARPES 
spectrum, matrix element effects\cite{Sep} notwithstanding. The peak of the 
spectral density function defined by taking momentum density cuts (MDCs), 
shown by yellow dots, follows the renormalized dispersion (orange dashed 
line) up to binding energy of about 200 meV.  It then disperses to higher 
energies rapidly (waterfall effect) as it catches up with the bare 
dispersion (red solid line) around $\Gamma$. In fact, near $\Gamma$, the dressed 
spectral peak lies slightly below the bare band. The width of the spectral 
function is largest in the intermediate energy range of 200-600~meV, where 
its slope also is the largest.  This behavior of the spectral function 
results from the presence of peaks in the real and imaginary parts of the 
self-energy in the 200-500~meV energy range discussed in connections with 
Figs. 1(a) and (b) above. It is also in accord with the waterfall effect 
observed in ARPES experiments, although the sharpness of the theoretically 
predicted waterfall in Fig. 1(c) is less severe than in experiments, which 
may be due to limitations of our model, including the approximations 
underlying our treatment of the susceptibility.

\begin{figure}  
            \resizebox{7.2cm}{!}{\includegraphics{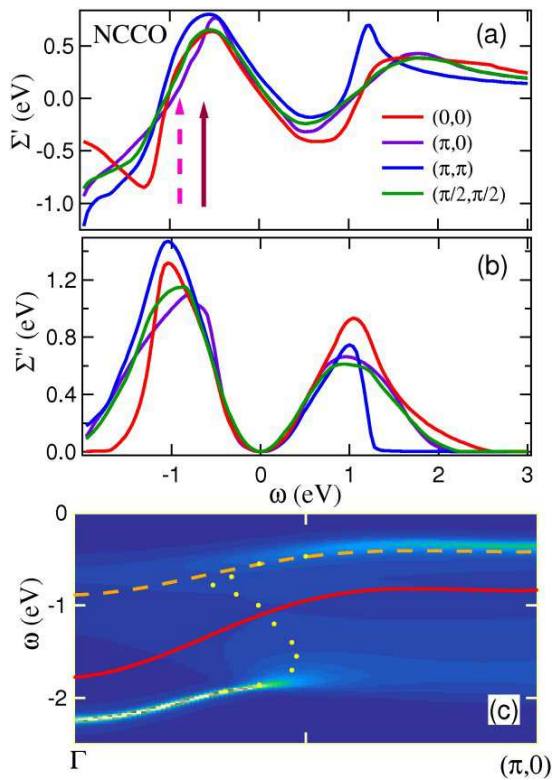}}
\vskip0.5cm
\caption{
(Color online) Same as the caption to Fig. 1, except that this figure 
shows only the computed results for overdoped NCCO. Experimental 
self-energies are 
not available and are therefore not shown (see text). Solid (dashed) arrow in (a)
shows location of peak in $\chi''$ at $(\pi ,0)$ ($(\pi /2,\pi /2)$).
}
\label{fig:10}
\end{figure}

Fig. 2 considers the case of electron doped (overdoped) NCCO. The peak 
in $\Sigma'$ in Fig. 2(a) lies at binding energies of 0.5-0.6~eV (in different $\vec 
k$-directions) with a height of 0.55-0.7~eV. Correspondingly, the peak in $\Sigma''$ in 
Fig. 2(b) lies at a binding energy of 0.7-1.1~eV with a height of 1-1.4~eV. Comparing 
these with the results of Fig. 1, we see that the self-energy effects in NCCO are much 
larger than in Bi2201. Our computed shift of $\sim$300~meV in the position of the 
peak in $\Sigma'$ to higher binding energy in going from Bi2201 to NCCO is 
in good accord with the experimentally reported shift of 
$\sim$300~meV\cite{PanDing}. The dispersion underlying the dressed Green 
function, which may be tracked through the yellow dots, is highly 
anomalous and presents a kink-like feature quite reminiscent of the more 
familiar low energy kinks in the 50~meV range around the $(\pi 
,0)$-direction\cite{LEK}, which have been discussed frequently in the 
cuprates. This strong bosonic coupling is also reflected in the fact that 
the band bottom in NCCO lies several hundred meVs below the bare LDA band 
in Fig. 2(c). It is interesting to note that the self-energies of Figs. 1 
and 2 display a `mirror-like' symmetry: The peaks below the Fermi energy 
in $\Sigma'$ and $\Sigma''$ for Bi2201 in Fig. 1 are smaller than those above 
the Fermi energy, but the situation reverses itself for NCCO in Fig. 2 in 
that now the peaks below the Fermi energy become larger than those above 
the Fermi energy.

\begin{figure}
            \resizebox{7.6cm}{!}{\includegraphics{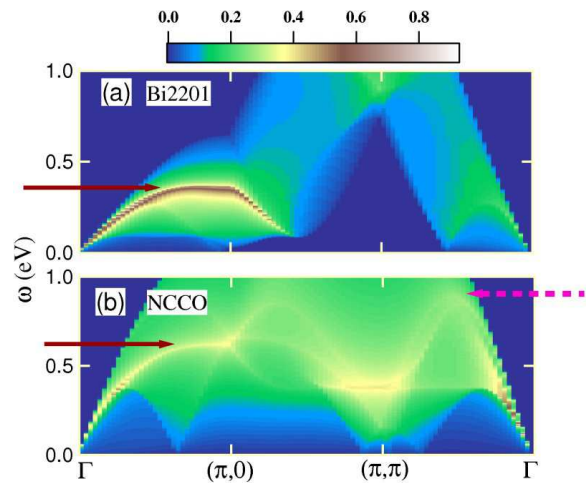}}
\vskip0.5cm
\caption{
(Color online) Map of the imaginary part of the magnetic susceptibility 
for (a) hole-doped Bi2201 and (b) electron-doped NCCO. Spectral weights 
are in units of $eV^{-1}$. Arrows mark the positions of the high spectral 
weights discussed in the text. 
}
\label{fig:7}
\end{figure}

The aforementioned shift of the peak in $\Sigma'$ to higher energy in NCCO 
can be understood in terms of the characteristics of the magnetic 
susceptibility.  Figure~~\ref{fig:7} compares in Bi2201 and NCCO the 
imaginary part $\chi''$, which is seen from Eq.~\ref{eq:4} to be related directly 
to the real as well as imaginary part of the self-energy. $\chi''$ is seen 
to be quite similar in shape along the $\Gamma$ to $(\pi ,0)$ line in 
Bi2201 and NCCO, except that in NCCO the band of high intensity (the 
yellowish trace) extends to a significantly higher energy scale. In 
contrast, $\chi''$ in the two systems differs sharply around $(\pi,\pi)$. 
These differences reflect those in the low-energy magnetic response of the 
two cuprates. NCCO with strong magnetic response around $(\pi,\pi)$ 
exhibits a nearly commensurate AFM order, while Bi2201 is very incommensurate,
with peaks shifted toward $(\pi ,0)$. In fact, the high energy peaks 
in the self-energy in Figs. 1 and 2 are tied to the flat-tops near $(\pi 
,0)$ at $\omega_1\sim 0.36$~eV in Bi2201 (solid arrow), and near both 
$(\pi ,0)$ at $\omega_2=0.62$~eV (solid arrow) and $(\pi /2,\pi /2)$ at 
$\omega_3 = 0.9$~eV (dashed arrow) in NCCO. Above these energies the weight 
in $\chi''$ falls rapidly, going to zero near an energy $8\bar t$. 

A reference to Figs. 1(a) and 2(a), where the energy $\omega_1$ in Bi2201, 
and the energies $\omega_2$ and $\omega_3$ in NCCO are marked by arrows, 
indicates that the peaks in $\Sigma'$ are correlated with these features in the 
magnetic susceptibility. In this spirit, the shift in the peak in $\Sigma'$ 
to higher energy in going from Bi2201 to NCCO reflects the fact that 
feature $\omega_3$ in $\chi''$ at $(\pi /2,\pi /2)$ in NCCO (dashed arrow 
in Fig.~2(a)) lies at a higher energy than the $(\pi ,0)$ feature 
$\omega_1$ in Bi2201 (arrow in Fig.~1(a)). Notably, when the Stoner factor 
$S=1/ (1-U\chi_0)$ is large, a peak in $\chi''$ arises from a peak in 
$\chi_0'(\omega )$, which in turn is associated with nesting of features 
separated by $\omega$ in energy.  In the present case, the nesting is from 
unoccupied states near the Van Hove singularity (VHS) at $(\pi ,0)$ to the 
vicinity of the band bottom at $\Gamma$, so $\omega_1\sim 2(t+2t')\sim 
0.32$~eV in Bi2201.  The larger value of $\omega_2$ in NCCO reflects the 
shift of the Fermi energy to higher energies in an electron-doped 
material.

A notable difference between electron and hole doping is the low-$\omega$ 
behavior of $\Sigma''$, which is quadratic in $\omega$ for electron-doping 
in Fig.~2(b), but nearly linear for hole-doping in Fig.~1(b).  The 
linearity for hole-doping, reminiscent of marginal Fermi liquid physics, 
is associated here with the proximity of the chemical potential to the 
VHS. This point is considered further in Fig.~4 
where $\Sigma''$ is shown in Bi2201 at the $(\pi,0)$ point for three 
different values of the chemical potential. When the chemical potential 
lies at the VHS (red line), $\Sigma''$ varies linearly, but when it is 
shifted by $75$ meV above or below the VHS, the behavior changes 
rapidly to become parabolic.
\begin{figure}  
            \resizebox{7.2cm}{!}{\includegraphics{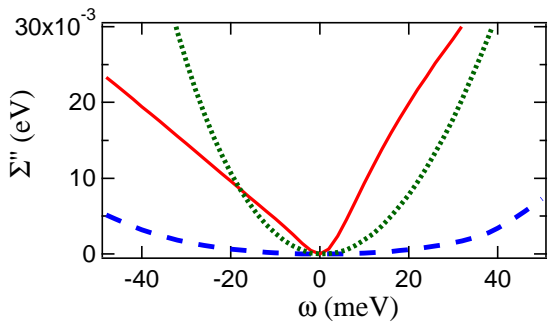}}
\vskip0.5cm
\caption{
(Color online) Low-energy behavior of $\Sigma ''(\omega )$ at $(\pi ,0)$ 
in Bi2201 for three different values of the chemical potential $\mu$ in 
relation to the position of the energy $E_{VHS}$ of the van Hove 
singularity. $E_{VHS}-\mu$=0 (red solid line); $E_{VHS}-\mu$= +$75$ meV 
(green dotted line); and $E_{VHS}-\mu$=--75 meV (blue dashed line).
} 
\label{fig:20}
\end{figure}

The strong magnetic scattering discussed in this study in the case of 
overdoped cuprates should persist into the underdoped regime, where the 
Stoner factor is expected to become larger.  In fact, this scattering is a 
{\it precursor} to the magnetically ordered state near half-filling and it 
is responsible for opening the magnetic gap. In contrast, a number of 
authors have related the presence of waterfall-like effects near 
half-filling to `Mott' physics associated with $(\pi ,\pi )$ AFM 
fluctuations\cite{KFul,Mano,WTW}, but have difficulty explaining why these 
effects persist into the overdoped regime.

The possible doping dependence of $U$ has been an important issue in 
connection with electron-doped cuprates.  A doping-dependent $U$ is 
suggested by a number of studies in the hole-doped cuprates as well. These 
include: Optical evidence of Mott gap decrease\cite{optU}; ARPES 
observation of very LDA-like bands in optimally and overdoped materials; 
models of the magnetic resonance peak\cite{magresU}; and, a strongly 
doping-dependent gap derived from Hall effect studies\cite{Ando}. The 
$\bar Z^2$-renormalization of $\chi_0$ in Eq.~\ref{eq:2} bears on this 
question and gives insight into how the value of $U$ enters into the 
magnetic response of the system. Recall that the susceptibility is often 
evaluated in the literature via Eq.~\ref{eq:3} using experimental band parameters, 
but without the $\bar Z^2$ factor of Eq.~\ref{eq:2} in $\chi_0$, which yields a 
$\chi$ scaling $\sim \bar Z^{-1}$ rather than the correct scaling of 
$\chi\sim\bar Z$. This can be corrected by replacing the $U$ in the Stoner 
factor by 
\begin{equation} U_{eff}=\bar Z^2U. 
\label{eq:5} 
\end{equation}
Indeed, our Hubbard parameter for NCCO of $U=4t$ is closer to the value at 
half-filling than is generally found.\cite{foot5}

In conclusion, we have shown that the higher energy magnetic 
susceptibility in the cuprates has considerable weight near $(\pi ,0)$ and 
that this leads to a high energy kink or waterfall-like effect in 
dispersion in both electron and hole-doped cuprates, providing an 
explanation of such effects observed recently in ARPES. Although our 
analysis is limited to the overdoped regime, we expect strong magnetic 
scattering to persist into the underdoped regime. This point however bears 
further study.

This work is supported by the US Department of Energy contract 
DE-AC03-76SF00098 and benefited from the allocation of supercomputer time 
at NERSC and Northeastern University's Advanced Scientific Computation 
Center (ASCC).


\begin{thebibliography}{99}
\bibitem{foot6}Although not discussed here, we find similar results in LSCO.
\bibitem{foot3}For Bi2201, we use the parameters of Bi$_2$Sr$_2 $CaCu$_2$O$_{8}$ 
(Bi2212), but neglect the bilayer splitting.  Following Ref.~\onlinecite{Arun3}, the hopping parameters 
are ($t,t',t'',t'''$) = (360,--100,35,10)~meV for Bi2201 and (420,--100,65,7.5)~meV for NCCO.
\bibitem{foot2}The specific values of $Z$ are 0.28 in Bi2212 and 0.55 in NCCO.\cite{Arun3}
\bibitem{foot4}Specifically, $\bar Z$ is the renormalization of the coherent part of the 
dispersion.
\bibitem{foot5}The smaller $U$ for Bi2201 may be due to neglect of $k_z$-dispersion, which 
broadens features in $\chi$ near the VHS.
\end{thebibliography}
\end{document}